\documentclass[
 preprint,
 amsmath,amssymb,
 aps, physrev,
floatfix,
]{revtex4-2}

\usepackage{graphicx}% Include figure files
\usepackage{dcolumn}% Align table columns on decimal point
\usepackage{bm}% bold math
\DeclareMathOperator{\arcsinh}{arcsinh}
\usepackage{float}
\begin{document}

%\preprint{APS/123-QED}

\title{\textbf{Physics of Dipole and Quadrupole Brewster Angles in Thin Films} 
}% 
\author{Edward H.Krock}
\email{ekrock@tcd.ie}
%\email{ekrock@tcd.ie}
 %\altaffiliation[Also at ]{Physics Department, XYZ University.}%Lines break automatically or can be forced with \\
\author{Mughaid Ali}%
\author{Haizhong Weng}%
\author{C. M. Smith}
\author{John F.Donegan}

\affiliation{
School of Physics, CRANN, and AMBER, Trinity College Dublin,
Dublin 2, Ireland
}

\date{\today}% It is always \today, today,
             %  but any date may be explicitly specified

\begin{abstract}

The Brewster angle is a well-known-phenomenon that describes the angle at which the intensity of reflection of p-polarized light is zero for a single dielectric interface. We investigate the angle dependent reflection in a simple SiN thin film, with thickness in the hundreds of nanometres, a common thickness used in optical waveguides, Fabry-Perot resonators, sensors and lasers. We describe the reflection of our SiN thin film in terms of electric and magnetic multipoles through a multipole expansion of the fields inside our film. Previous theoretical studies on Fabry Perot modes in GaP films have only considered the reflection of unpolarized light at normal incidence. Our investigation expands on this work to s- and p-polarization and angle dependent effects permitting the study of both Fabry Perot and Brewster angle effects together. Our approach allows us to re-derive the well-known Brewster angle equation from the electric dipole term. We then derive several new Brewster angle equations associated with the magnetic dipole and electric/magnetic quadrupoles in our model. Our model is then validated by obtaining good agreement between the predicted reflection from our multipoles to the measured reflection of the same thin film. The distinction between the standard electric dipole Brewster angle and our newly discovered Brewster angles is the destructive interference between remaining multipoles. It is this destructive interference which produces the zero in measured and modelled reflection, associated with the Brewster angle. In addition, the Brewster condition of the magnetic dipole and quadrupoles are only satisfied at specific wavelengths. This multipole model brings additional understanding of how light interacts with thin film dielectric materials.
\end{abstract}

%\keywords{Suggested keywords}%Use showkeys class option if keyword
                              %display desired
\maketitle

%\tableofcontents
\newpage
\section{Introduction}\label{sec1}

In 1815 \cite{DavidBrewster} David Brewster observed a simple law for the angle of incidence, $\theta_{B}$, at which an unpolarized incident beam is perfectly s-polarized upon reflection from a dielectric medium with refractive index $n$. 
\begin{equation}
    \tan(\theta_{B})=n
\end{equation}

\begin{figure}[H]
\centering
\includegraphics[width=0.9\textwidth]{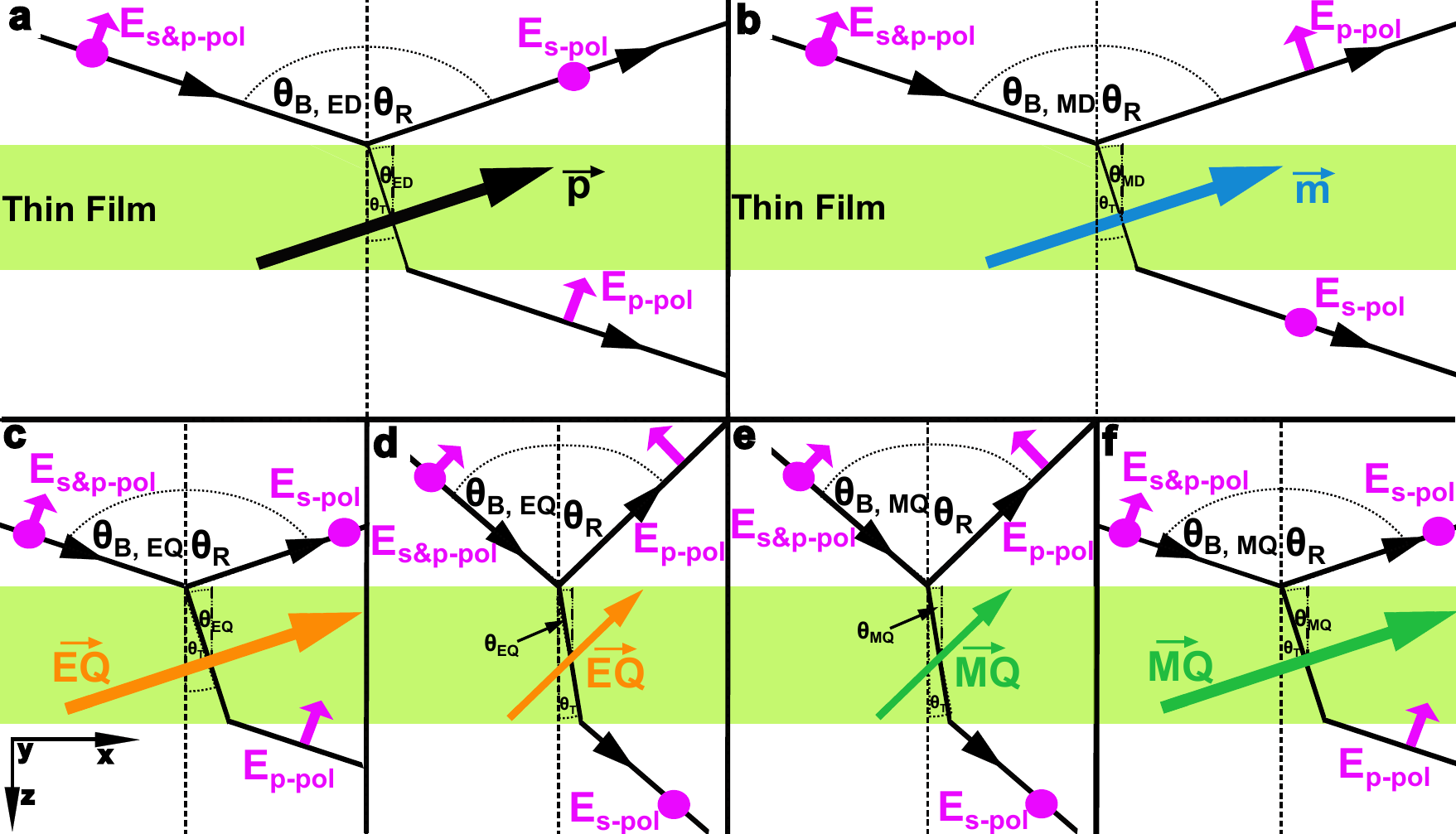}
\caption{Illustration of geometry for incident (s- and p- polarized), reflected and transmitted electric fields from a thin film considering each multipole on its own. Subplot (a) illustrates the traditional electric dipole Brewster angle, the electric dipole moment, $\vec{p}$, aligns with the reflected direction,$\vec{n}$, and the reflected light is perfectly s-polarized. However, the Brewster angle can be observed for all multipoles as we show in this work, as any multipole moment may align with the reflected ray. Subplots (b) illustrate this for the magnetic dipole, where the reflected ray is perfectly p-polarized. Subplots (c)-(d) illustrate this for the electric/magnetic quadrupoles, where the reflected ray may be perfectly s- or p-polarized depending on the direction of the incident ray.}
\label{fig1}
\end{figure}

This leads to the current understanding of the Brewster angle illustrated in Figure \ref{fig1} (a). The radiation pattern of an electric dipole has a minimum along the direction of the dipole moment. If this dipole moment aligns with the direction of reflection then no light is reflected \cite{kong1990}. This is equivalent to the angle of incidence/reflection being equal to the angle of the dipole moment. 
In more modern-day usage, a zero in reflected intensity is generally considered to occur at the Brewster angle \cite{Lakhtakia:89}. In the past decade, this picture has been further broadened to what is termed a generalized Brewster angle effect \cite{Sreekanth2019,Paniagua2016,2021GBE,Tiukuvaara23}. In these studies, the behaviour of the electric dipole moment is not studied directly, but a zero reflection/polarizing angle is identified. Through the use of metamaterials or optical absorbing layers, the generalized Brewster angle effect can be observed for both s-and p-polarized incident light. Thus, zero reflection can be engineered at multiple wavelengths and at multiple angles of incidence. To this day, Brewster angles are still a topic of continued research with 2D materials \cite{Yermakov24,Majrus_2018} and the observation of the anomalous spin-Hall effect  near the Brewster angle \cite{Ling21}.

We study the Brewster angle by describing the reflection of light from a simple dielectric material in terms of electric and magnetic multipoles, known as Multipole Decomposition. While such descriptions are discussed in textbooks \cite{jackson_classical_1999}, direct application to simple uniform thin films is unusual. Optical metamaterials \cite{Evlyukhin2012,Majorel2022} are more usually investigated. Multipole decomposition is vital in the understanding of a variety of metamaterial phenomena, including toroidal multipoles \cite{Kaelberer2010}, which play a role in the observation of anapoles \cite{Savinov2019,Baryshnikova2019,Miroshnichenko2015}. Other interference phenomena such as Kerker effects, the interference between electric and magnetic multipoles \cite{Terekhov2019,Alaee2020,Bukhari2021} also occur with metamaterials. The presence of magnetic multipoles at frequencies where the bulk permeability of materials is unity \cite{Merlin2009,Monticone2014} is known as ”Optical Magnetism”, \cite{Monticone2014,Evlyukhin2012,Shafiei2013} which enables metamaterials to support unusual phenomena such as negative refraction \cite{Pendry2004} and transverse electric surface waves \cite{Papadakis2018}.

To our knowledge, only a single multipole decomposition study on  the reflection of thin films has been performed to date, which involved a subwavelength film of 200 nm GaP studied at normal incidence and unpolarized light \cite{Li2021}. They demonstrated that Fabry Perot modes arise from the interference between electric and magnetic multipoles, highlighting that "Optical Magnetism" is not unique to metamaterials, but an intrinsic property of thin films. However, combined experimental evidence and theoretical interpretation of polarized and angle dependent behaviour has not been explored, and is the focus of this work. Thin films are a fundamental component of modern optics and are used in a wide variety of applications in lasers, waveguides \cite{Blumenthal,Zhu:21,Jalali07}, optical absorbers \cite{Kats}, colour control \cite{Daqiqeh21} and sensors. Hence studying the reflection of such films with a more fundamental model is valuable to develop a better understanding of these systems.\par

From multipole decomposition, we derive conditions for when reflection is zero for individual multipoles. For the electric dipole, we find that this condition is the well-known Brewster condition, Figure 1 (a), that can be used to derive the well-known  Brewster angle equation, equation 1. The other conditions derived for the magnetic dipole and electric/magnetic quadrupoles are also Brewster angle conditions of their own, which lead to new, previously unidentified Brewster angle equations of their own. These new Brewster angles for these multipoles as illustrated in Figure 1 (b)-(d) where the respective multipole moments are shown to align with the direction of reflection.\par

For the Brewster angle in thin films, we find that the alignment of the multipole moment with the reflection direction must be considered in the complex plane as our multipole analysis produces complex angles. That is, the Brewster angle condition should consider complex angles. 
\par 

We also observe zeros in reflection resulting from the presence of Fabry Perot modes at specific wavelengths and angles, but unlike the Brewster angles, they occur for both s- and p-polarized light. In our model, these Fabry Perot modes can be interpreted as interference between electric and magnetic multipoles without the alignment between multipoles and the reflection direction. This result agrees with similar theoretical studies at normal incidence,\cite{Li2021}, however we extend this result to all angles of incidence.\par

Our work uses a more complex model to describe the reflection of light to broaden and complete the understanding of the Brewster angle in thin films of well-defined optical thicknesses of about 500 nm. We highlight the significant role that higher order electric and magnetic multipoles play in the observation of Brewster angles for these thin films.\par

\section{Multipole Theory of Brewster Angles}\label{sec2}

Multipole decomposition describes the complex reflection coefficient,$r$, from a dielectric thin film in terms of electric and magnetic multipoles with equation 2 \cite{Evlyukhin2019}.\par

These multipole moments are shown in the expected order of their contribution to the overall reflection, which are the electric dipole (ED), magnetic dipole (MD), electric quadruple (EQ), magnetic quadrupole (MQ), electric octupole (EO) and magnetic octupoles (MO). Our model extends this behaviour to all angles of incidence, implementation and derivation of this model is explained in the supplemental material, section I.\par

\begin{multline}    
r = \frac{i k_{d} }{E_{0}2S_{L} \epsilon_{0} \epsilon_{d}} [\vec{n} \times (\vec{p} \times \vec{n})+ \frac{1}{v_{d}}[\vec{m} \times \vec{n}]+\frac{ik_{d}}{6}[\vec{n}\times (\vec{n}\times \hat{Q}\vec{n})]+\\\frac{ik_{d}}{2v_{d}}[\vec{n} \times \vec{M}\vec{n}]
+\frac{k_{d}^{2}}{6}[\vec{n} \times [\vec{n} \times \hat{O^{e}}\vec{n}\vec{n}]] +\frac{k_{d}^{2}}{6 v_{d}} [\vec{n} \times \hat{O^{m}}\vec{n}\vec{n}]]%\vec{n}]\\+
\end{multline}

Figure \ref{fig1} illustrates our Brewster angles and includes several additional angles, $\theta_{B}$ describes the angle of incidence that the Brewster angle occurs at.  Angles of incidence $\theta_{I}$, transmission $\theta_{T}$ and angle of reflection $\theta_{R}$ are well-known . Each multipole will have its own Brewster angle which we denote for electric dipole $\theta_{B,ED}$. In addition, each multipole has its own angle $\theta_{ED}, \theta_{MD}, \theta_{EQ}, \theta_{MQ}$ which are complex numbers. This is due to the complex valued electric fields used to calculate the multipoles moments. In the case of the electric field, the imaginary part describes the time varying part of the electric field and for multipoles, indicates an oscillating multipole that is emitting light\cite{jackson_classical_1999}.

In our analysis of reflection and transmission of light from a thin film, we begin by considering each of the terms in equation 2 on their own. Later we will look at their combined effects in the model and in experiment. \par

\subsection{Electric Dipole Brewster Angle}
We first derive the Brewster angle arising from the electric dipole term. In equation 2, the resulting cross products between $\vec{n}$ and $\vec{p}$ leads to the following vector for the reflected light. The zero-reflection case which we expect at the Brewster angle should correspond to when $\vec{n}$ and $\vec{p}$  are parallel. We note that $\vec{p}$ is typically a complex vector given the electric field is itself complex and hence the multipole angles calculated in the multipole analysis are complex in general, but we will show that at the Brewster angle, the complex electric dipole moment becomes a real vector. From the first term in equation 2, $\vec{n} \times (\vec{p} \times \vec{n})$, we have the following vector for reflection from an electric dipole. \par
\begin{equation}    
\vec{n} \times (\vec{p} \times \vec{n})=\begin{bmatrix} \cos( \theta_{I} )(p_{x}\cos( \theta_{I} )-p_{z}\sin( \theta_{I} )) \\ p_{y} \\ \ \sin(\theta_{I}) (p_{z}\sin( \theta_{I} )-p_{x}\cos( \theta_{I} ))   \end{bmatrix}
\end{equation}

The first/third element of this vector describes p-polarized light, the second element describes s-polarized light, and has no angle dependence. Therefore, the electric dipole supports a single p-polarized Brewster angle as $p_{y}$ can never be zero. Setting the first element equal to zero gives us equation 4.
\begin{equation}
    \cos( \theta_{I} )(p_{x}\cos( \theta_{I} )-p_{z}\sin( \theta_{I} ))=0
\end{equation}

Making the angle of incidence equal to the Brewster angle, $\theta_{I}=\theta_{B,ED}$ and simplifying equation 4 results in the following condition.  
\begin{equation}
    \tan(\theta_{B,ED})=\frac{p_{x}}{p_{z}}
\end{equation}

Equation 5 describes a condition where the electric dipole does not reflect any light, which occurs when the tangent of the angle of incidence is equal to the ratio of the complex electric dipole moment along the x- and z-directions. In order for equation 5 to be satisfied, and for the Brewster angle to occur, these moments must be real numbers. Next, we show that equation 5 is the Brewster angle equation 1 by deriving the Brewster condition, $\theta_{i}+\theta_{t}=90^{\circ}$, from equation 5. 

The angle of the complex electric dipole moment is the inverse tangent of the electric dipole moment ratio in equation 5.
\begin{equation}    
atan\left( \frac{p_{x}}{p_{z}}\right)=\tilde\theta_{ED}\\=\theta_{ED,r}+i\theta_{ED,i}
\end{equation}

Applying the inverse tangent to equation 5, gives us a modified form of the Brewster angle condition.
\begin{equation}
    \theta_{ED,r}-\theta_{I}+i\theta_{ED,i}=0^{\circ}+0i
\end{equation}

$\theta_{ED,r}$ and $\theta_{ED,i}$ are the real and imaginary parts of the electric dipole angle. We obtain the Brewster angle condition in terms of the angle of incidence/reflection and angle of transmission, by realizing the angle of transmission is related to $\theta_{ED,real}$ via a $90^{\circ}$ rotation, as the direction of propagation and electric field are perpendicular to each other, $\theta_{t}=\theta_{ED,real}-90^{\circ}$, as we are dealing with transverse waves.\par

\begin{equation}
   \theta_{i}+\theta_{t}=90^{\circ}
\end{equation}

We now can derive equation 1 from Snell's Law, $\sin(\theta_{i})=n\sin(\theta_{t})$, using equation 8. This derivation demonstrates that the zero reflection from the electric dipole term corresponds to the Brewster angle defined in equation 1.

\begin{equation}
    \tan(\theta_{B,ED})=\frac{p_{x}}{p_{z}}=n
\end{equation}
Equation 7 and 9 highlights that at the Brewster angle, the dipole moment should be a real number, as shown by the imaginary part of equation 7, $\theta_{ED,i}=0$ and the ratio of the dipole moments in equation 9, $\frac{p_{x}}{p_{z}}$, being equal to the refractive index, which is a real number.  Later, we will show that $\theta_{ED,i}=0$ at the ED Brewster angle and we show in the supplemental material, section XI, that the ratio, $\frac{p_{x}}{p_{z}}$, is real at the Brewster angle.

This is an expected result as the reflection coefficient is defined by a cross product between $\vec{n}$ and $\vec{p}$, $\vec{n}$ is a unit vector which is purely real, while $\vec{p}$ is a complex vector. To obtain zero reflection from a cross product, $\vec{n}$ and $\vec{p}$ must be parallel, which can only happen if $\vec{p}$ is real. A complex dipole moment means that the dipole is reflecting light \cite{jackson_classical_1999}, which does not occur at the Brewster angle. The real valued dipole moment in our derivation is demonstrating that the ED is not reflecting light at the Brewster angle. In our case, the imaginary part is non-zero at every angle, except at the Brewster angle, which we show later. When we apply our model to an absorbing material, $n+ik=2+2i$ (where k is the extinction coefficient), in the supplemental material, section X, we find that the imaginary part is never zero, which is correct for absorbing scatterers according to the Optical Theorem \cite{jackson_classical_1999} and the reflection is never zero. In fact, the ratio of the imaginary parts is equal to the extinction coefficient k.\par

%Another possibility is that $\vec{n}$ is a complex vector that aligns a complex valued $\vec{p}$, this may occur if the angle of incidence is a complex number,  which is not possible in our system. 

%Optical Theorem \cite{jackson_classical_1999}suggest that the imaginary part of the scattering amplitude, which is related to our dipole moment, can never be zero, but this is a mistake in interpreting Optical Theorem. When considering non-absorbing systems, one must consider all angles, as considering a single angle may give zero. 

\subsection{Magnetic Dipole Brewster Angle}
We can now derive the angle of incidence that satisfies the Brewster angle condition for any multipole using the approach outlined in the previous section. As with the electric dipole derivation, we presume that only a single multipole is contributing to reflection. The detailed derivations are available in the supplemental material, section I, and we only present the derived equations here. For the magnetic dipole we obtain a single s-polarized condition where $m_{x}$ is the magnetic dipole moment along the x-direction, $m_{z}$ is the magnetic dipole moment along the z-direction.
\begin{equation}
\tan(\theta_{B,MD,s-pol})=\frac{m_{x}}{m_{z}}
\end{equation}
While the ED Brewster angle only occurs for p-polarized light, the MD Brewster angle follows a similar requirement that the complex dipole moment must be a real number.

\subsection{Quadrupole Brewster Angles}

For quadrupoles, we obtain a condition for each polarization. $EQ_{ab}$ and $MQ_{ab}$ are the electric/magnetic quadrupoles moments in the a-b plane (or direction if a=b), where $a$ and $b$ are axes in the cartesian plane, $x,y,z$, see Figure \ref{fig1}.

\begin{equation}    
    \tan(2\theta_{B,EQ,p-pol})=\\-\frac{2EQ_{xz}}{EQ_{zz}-EQ_{xx}}
\end{equation}

\begin{equation}
    \tan(\theta_{B,MQ,p-pol})=\frac{MQ_{zy}}{MQ_{xy}}
\end{equation}

\begin{equation}
    \tan(\theta_{B,EQ,s-pol})=\frac{EQ_{zy}}{EQ_{xy}}
\end{equation}

\begin{equation} 
    \tan(2\theta_{B,MQ,s-pol})=\\-\frac{2MQ_{xz}}{MQ_{zz}-MQ_{xx}}
\end{equation}

Unlike their dipole counterparts, the quadrupole Brewster angles can occur for both polarizations at different angles. These new Brewster angles are shown schematically in Figure \ref{fig1} (c) and (d), where two Brewster angles occur for each multipole.\par

\subsection{Satisfying the Brewster Condition for Each Individual Multipole}
\begin{figure}[H]
\includegraphics[width=0.95\columnwidth]{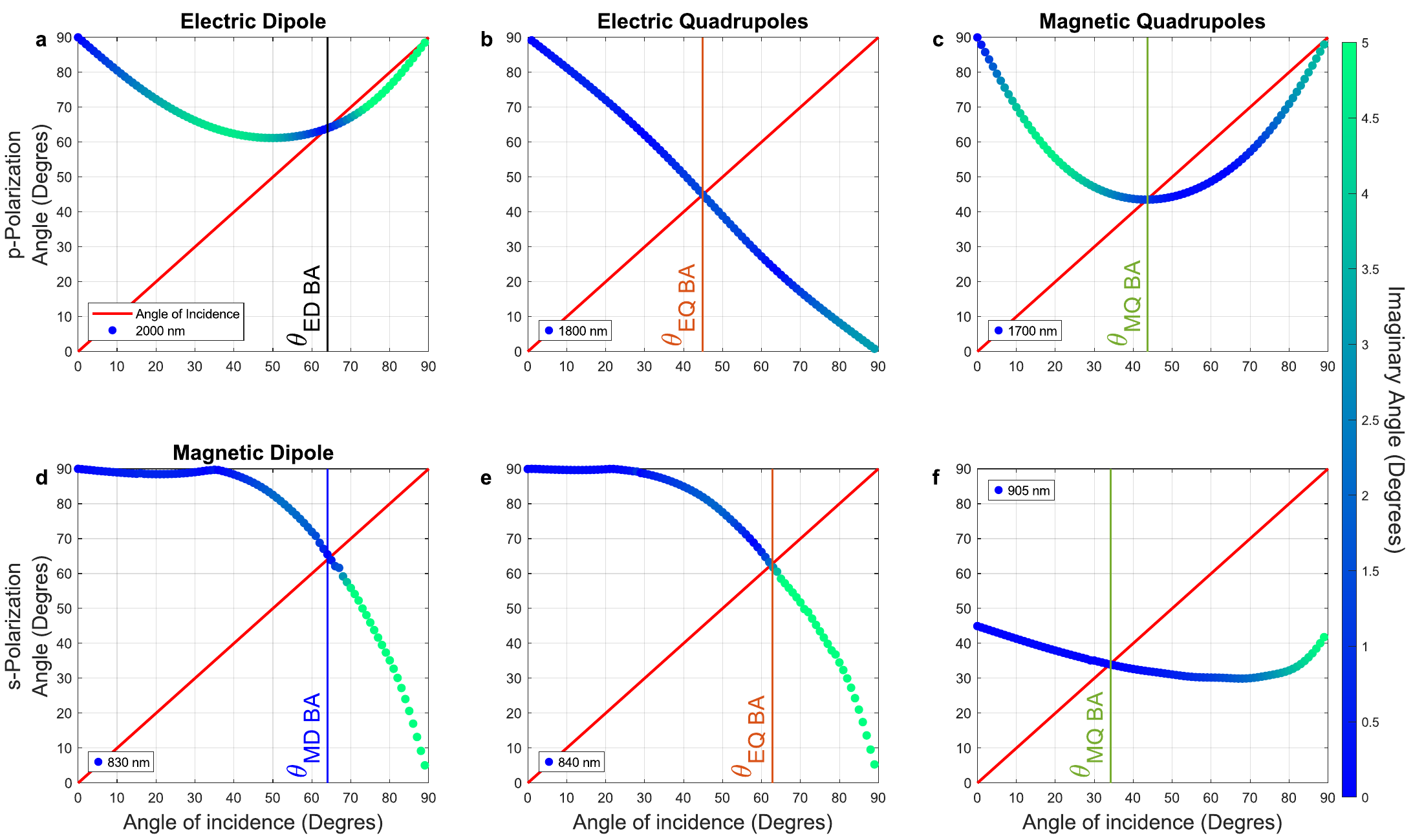}
\label{fig2.1}
\caption{Demonstration of the Brewster angle condition, $\theta_{Multipole,real}-\theta_{I}+i\theta_{Multipole,imag}=0^{\circ}+0i$  being satisfied for p-polarized reflection for the (a) electric dipole, (b) electric quadrupole, (c) magnetic quadrupole and s-polarized reflection, (d) magnetic dipole, (e) electric quadrupole, (f) magnetic quadrupole. The Brewster angle condition is satisfied when the real part of the angle of the multipole aligns with the angle of incidence, as marked by the intersection point with the red line, as well as the imaginary part of the angle of the multipole being zero, marked by the dark blue colour. The colour represents the imaginary angle, which is restricted to values between 0-5 for visual clarity, but the value may be larger than 5. Values greater than 5 correspond to the brightest green colours. }
\end{figure}

To demonstrate equations 5 and 8-12, we study a 455 nm thick film of free-standing SiN, with a dispersive refractive index detailed in the supplemental material, section VI. SiN is a dielectric material with a bandgap in the UV and a refractive index near 2. The expected electric dipole Brewster angle calculated from equation 1 is roughly 64 degrees corresponding to a refractive index of 2.0. We calculate the inverse tangent for equations 5 and 8-12  and show in Figure 2 where they intersect with the angle of incidence, which represents the Brewster angle condition in equation 7 being satisfied for each individual multipole. The y-axis plots the real part of the complex angle, while the scatter plot and colour scale indicate the imaginary part of the angle. The imaginary part of the angle can be related to the magnitude of the imaginary part of $\vec{p}$ through the relation $\theta_{ED,i}=\arcsinh({\frac{|imag(\vec{p})|}{|(\vec{p})|}})$, as discussed in supplementary material, section II.

In Figure 2 (a), the p-polarized electric dipole angle is plotted, and it is shown that the real part of the angle intersects with the angle of incidence at the expected position of the Brewster angle at roughly 64 degrees. At the same time, the dark blue colour shows where the imaginary part is zero and which is now in addition to the Brewster angle condition in equation 5. The p-polarized EQ and MQ angles are plotted in Figure 2 (b)-(c). For the p-polarized EQ Brewster angle in Figure 2 (b) this is roughly 45 degrees at a wavelength of 1800 nm. For the p-polarized MQ Brewster angle in Figure 2 (c) this is roughly at 42 degrees at a wavelength of 1700 nm. For the s-polarized Brewster angles in Figure 2 (d)-(e), we observe that the MD and EQ Brewster angles occur at the same angle of roughly 64 degrees as the ED case (at 830 nm). We can now say that the well-known Brewster angle equation applies to p-polarized light for the ED case and s-polarized incident light for the MD and EQ cases as shown in equation 13. The wavelengths chosen are from a more detailed wavelength map that will be described later.

It is possible to demonstrate that the magnetic dipole Brewster angle, equation 8, leads to the same tangent equation as the electric dipole Brewster angle in equation 5, as is shown in supplemental material, section II D. Assuming the second term of the EQ moment is sufficiently small, then the behaviour of the s-polarization EQ Brewster angle is the same as that of the MD, which we demonstrate in the supplemental material, section II E. We also show in Figures S9 (a) and (d) and S10 (a)-(e) that the fractions presented in equations 9, 10 and are all equal to 2 at their Brewster angles of 64 degrees. A similar equivalence has been demonstrated in the reflection coefficients of these two multipoles for normal incidence elsewhere \cite{Li2021}. Thus, we can state that the ED, MD and EQ Brewster angle occur at the same angle, but different polarizations and wavelengths.\par

\begin{equation}
\tan(\theta_{B}) =\frac{p_{x}}{p_{z}}=\frac{m_{x}}{m_{z}}=\frac{EQ_{zy}}{EQ_{xy}}=n =2
\end{equation}

\section{Experiment and simulation of reflection from a thin-film of silicon nitride}\label{sec2}
\subsection{Angle Dependent Reflection at 1500 nm}
\begin{figure}[b!]
\includegraphics[width=1\textwidth]{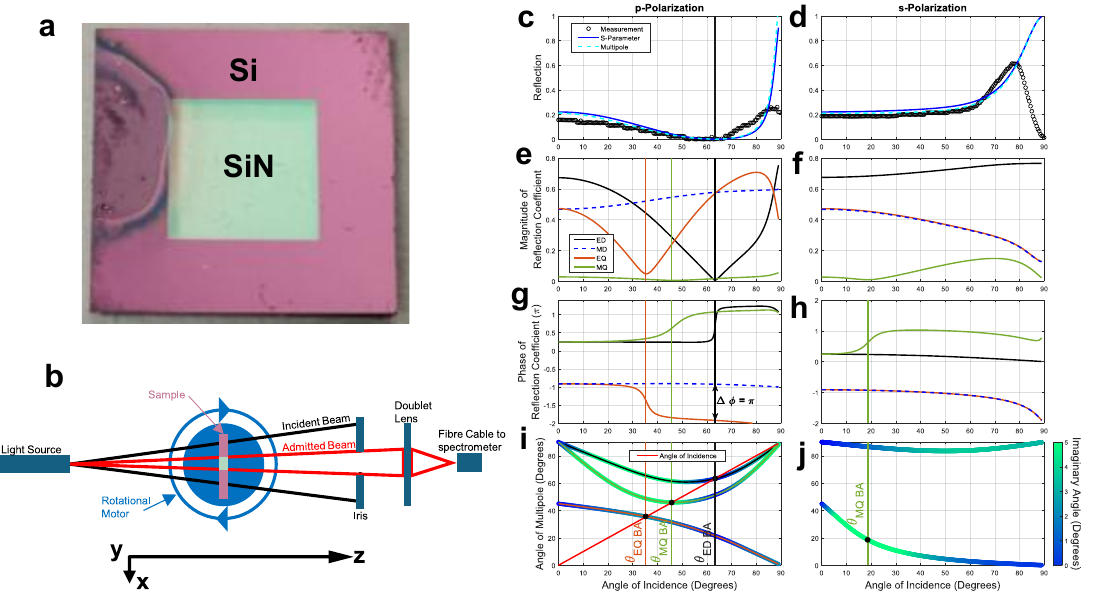}
\caption{Comparison of our multipole model with the S-parameter simulation and experiment via a study of (a) free standing SiN thin film of thickness 455 nm at a wavelength of 1500 nm. (b) Schematic of experiment to measure reflection. (c)-(d) Comparison between measured (circles) and modelled reflection with S-parameter (blue line) and multipole decomposition (dashed cyan line) for both (c) s-polarization  and (d) p-polarization . (e)-(f) Amplitude of individual multipole reflection coefficients, ED (black line), MD (dashed blue line), EQ (orange line), MQ (green line) in the multipole model for both (e) s-polarization  and (f) p-polarization . (g)-(h) Unwrapped phase of individual multipole reflection coefficients in the multipole model, phases are divided by $\pi$.(i)-(j) Angle of multipoles as previously shown in Figure 2 for both (i) s-polarization  and (j) p-polarization. Vertical black, green and orange lines mark the ED, MQ and EQ Brewster angles.}
\label{fig2}
\end{figure}
While we have derived new Brewster angles associated with each individual multipole, in order to describe the reflection of a thin film, we need to consider all multipoles together as described in equation 2, as equations 5, 8-12 only identify the angles at which the individual Brewster angle occurs. It has been previously demonstrated that the reflection spectra, at normal incidence, of a thin film is composed of many multipoles interfering with each other \cite{Li2021} producing cases of zero reflection in a Fabry Perot arrangement.\par

We study how our model predicts the angle dependent reflection spectra of a SiN thin film at 1500 nm. SiN is fast becoming the key material platform for much of integrated photonics from comb sources to single photon emitters \cite{HosseinEnjavi:25}. We investigate through experiment and theory a free-standing 455 nm thin film of SiN supported by a Si frame at a single wavelength of 1500 nm in Figure \ref{fig2} (a) and over a wavelength range from 750-2200 nm in Figure \ref{figReflectionBroad}. The experimental set up is shown in Figure \ref{fig2} (b). 

Figure \ref{fig2} (c)-(d), verifies equation 2, which is our multipole model as a whole, where we describe the reflection of our SiN thin film in terms of the combined effects of the electric and magnetic multipoles. There is excellent agreement between the multipole and S-parameter models over the angles and wavelength studied. The S-parameter model is used extensively in modelling reflection and transmission of light \cite{COMSOLGuide}. It is striking that the agreement with our multipole model is so good. Clearly, reflection from thin films in better understood with the multipole model in which we can see the contributions of the individual terms.\par  

\subsubsection{Role of Destructive Interference in the ED Brewster Angle}

In Figure \ref{fig2} (e)-(f) we plot the magnitude of the individual reflection coefficients for the ED, MD, EQ and MQ terms for p- and s- polarized incident light as a function of angle of incidence. We neglect contributions from the octopoles terms in equation 2. Figure \ref{fig2} (g)-(h) shows the relative unwrapped phase (with respect to the phase of the incident light) of the various multipoles, where 2$\pi$ discontinuities resulting from the tangent function have been removed. Figure \ref{fig2} (i)-(j) shows the angle of the multipoles, highlighting the angle at which the Brewster angle condition is satisfied. \par

We can identify the positions of the ED Brewster angle from the reflection minima in from Figure \ref{fig2} (e), or the intersection marked in Figure \ref{fig2} (i). At 64 degrees, the contribution of the ED term is indeed zero as expected, $|r_{ED,~p-pol}| = 0$. We note that at the Brewster angle, the phase of the electric dipole coefficient in Figure \ref{fig2} (g) undergoes a $\pi$ phase shift, and this serves as another way to identify the Brewster angle. This $\pi$ phase shift corresponds to the imaginary part of the dipole moment becoming zero as we discussed earlier. We note the new result that the MD and EQ terms are large and equal in  Figure \ref{fig2} (e) but according to Figure \ref{fig2} (g) are out of phase by $\pi$ so that $|r_{MD,~p-pol}| + |r_{EQ,~p-pol}| = 0$, maintaining the well-known zero in reflection at this angle.\par
\subsubsection{Presence of Quadrupole Brewster Angles at 1500 nm}
There is a p-polarized MQ Brewster angle at 46 degrees and an EQ Brewster angle at 35 degrees in Figure \ref{fig2} (e). These are incomplete Brewster angles, as the reflection minima are not perfectly zero as seen in Figure \ref{fig2} (e) and the imaginary angles are non-zero in Figure \ref{fig2} (i). These Brewster angles are wavelength specific, and the studied wavelength is close to the exact wavelength required to observe these Brewster angles. This wavelength dependent behaviour is discussed later in Figure 4 (e), (i) and (j). Only a single s-polarized MQ Brewster angle is found in this case at 19 degrees in Figure \ref{fig2} (d)-(j). This is incomplete at this wavelength, due to the non-zero imaginary angle and due to incomplete cancellation from the other multipoles, a zero in reflection (Figure \ref{fig2} (d)) is not  found. \par

\subsection{Angle Dependent Reflection from 400-2200 nm}

\begin{figure}[h!]
\centering
\includegraphics[width=1\textwidth]{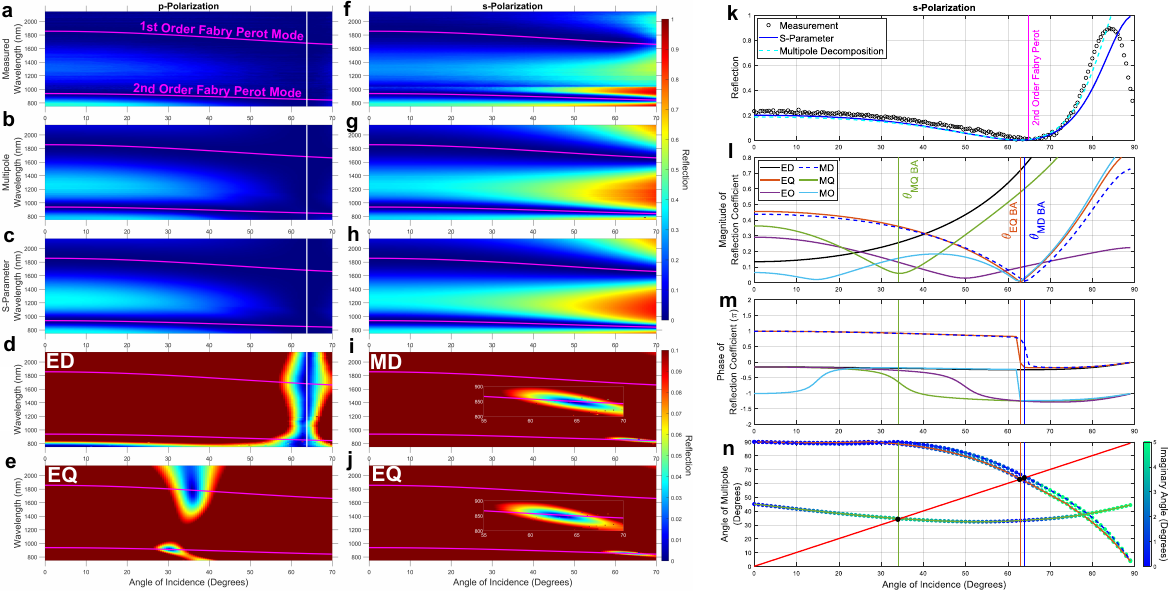}
\caption{Comparison between simulations and experiment from wavelengths 750-2200 nm. (a) and (f) measured reflection, (b) and (g) multipole reflection, (c) and (h) S-parameter reflection for both s- and p-polarizations, respectively. Magenta lines represent the first 2 Fabry Perot modes plotted by equation S.74. (d)-(j) Individual reflection coefficients for  (d) ED, (e) MD,  and (j) EQ for p- and s-polarization over a limited range of reflection from 0 to 0.1 to identify the Brewster angles. (k)-(n) Investigation of reflection at 850 nm, where the MD and EQ BAs are present similar to Figure \ref{fig2}. (k) Comparison between measured and modelled reflection. (l) Magnitude of individual reflection coefficients. (m) Phase of individual reflection coefficients. (n) Angle of MD,EQ and MQ multipoles.}
\label{fig3}
\label{figReflectionBroad}
\end{figure}
To fully explore the agreement between experiment and model, we plot the reflection spectra from 750-2200 nm and 0-70 degrees angle of incidence for p- and s-polarized light. These maps are shown in Figure \ref{figReflectionBroad} (a)-(c) and (f)-(h) and again show the excellent agreement between the multipole model and the S-parameter model as well as good agreement with the experimental data.
We see the clear presence of the 1st and 2nd Fabry Perot modes that will also result in zero reflection for both polarizations. Several regions of zero reflection match perfectly with the expected location of the Fabry Perot modes, whose location is calculated with equation S.74 in the supplementary information and marked with the magenta lines. The white line marks the location of the ED Brewster angle according to equation 1. The multipole model in Figure \ref{figReflectionBroad} (b) and (g) begins to break down at larger angles and shorter wavelengths, given that higher order multipoles in the model are required as the film thickness approaches the wavelength of the incident light \cite{Li2021}. For wavelengths in the visible we require octupoles to accurately predict reflection.\par

\subsubsection{Wavelength Specific Nature of MQ,EQ and MQ Brewster Angles}
As highlighted earlier, the new Brewster angles are wavelength dependent, and this is shown clearly in Figures \ref{figReflectionBroad} (e), (i) and (j). Here we plot the reflection coefficients for each multipole but over a limited range from 0 to 0.1 so that Brewster angle minima can be clearly seen. We contrast this to the individual reflection for the ED in Figure \ref{figReflectionBroad} (d), the ED Brewster angle can be seen at all wavelengths near 64 degrees while the EQ BA in Figure \ref{figReflectionBroad} (e) occurs near 35 degrees in the wavelength range from 1400-2100 nm as well as at 925 nm. In Figure \ref{figReflectionBroad} (i)-(j), we first note that the MD and EQ maps are identical as we met earlier. These Brewster angle occurs at 64 degrees in agreement with equation 13, but most striking is the very narrow range of wavelengths ($\sim$ 30 nm) over which it occurs in stark contrast to the ED case.\par
\subsubsection{Separating Fabry Perot Modes and Brewster Angles}
Examining the reflection from both experiment and simulations at 850 nm for s-polarized as a function of angle, we plot the results in Figure \ref{figReflectionBroad} (k),(l),(m),(n). We note in Figure \ref{figReflectionBroad} (k), the reflection is zero around 64 degrees, but this position also coincides with a Fabry Perot resonance, which happens to intersect with these Brewster angles. It is easy to confuse both phenomena, as they appear as a reflection minimum at a single angle. The difference between Fabry Perot and Brewster angles can be seen by changing polarization, the Fabry Perot mode is present for both polarizations, as seen in Figure \ref{figReflectionBroad} (a) and (f), while the Brewster angle is polarization specific, as shown in Figure \ref{figReflectionBroad} (a), (d), (e),(i) and (j).\par

In supplemental material, section IX, Fabry Perot modes and Brewster angles are distinctly different interference phenomena which happen to overlap at this wavelength and angle. In Figure \ref{figReflectionBroad} (l), we see the MD and EQ Brewster angles minima, but we note also the ED, MQ and EO terms are non-zero but their destructive interference produces the zero reflection minimum. The MQ is an incomplete Brewster angle, given the reflection minima and imaginary angle are non-zero. %There appears to be evidence of octupole Brewster angles in Figure \ref{figReflectionBroad} (l). The magnetic octupole shows two minima at 15 and 64 degrees respectively and the electric octupole shows a single minimum at 40 degrees.\par

\section{Conclusion}\label{sec13}

We have demonstrated a fundamental analysis of the Brewster angle in SiN thin films, showing how it arises from the electric dipole response of a medium, as it is traditionally understood with verification in an experiment with a free-standing SiN thin film. Our approach reveals a previously hidden feature of the Brewster angle in thin films. The destructive interference of electric and magnetic multipoles is essential to obtaining a zero reflection minimum. This work also highlights how other multipoles, including the magnetic dipole and electric/magnetic quadrupoles may also satisfy the well-known Brewster angle condition. What distinguished the electric dipole Brewster angle from our new magnetic dipole and quadrupole Brewster angles is the destructive interference between remaining multipoles.\par

The pseudo-Brewster angle which shows a non-zero reflection minimum in p-polarization for lossy materials can also be accounted for with our model. The non-zero reflection has been demonstrated to occur due to be a lack of alignment of the electric dipole with the angle of incidence, as well as an angular shift in the destructive interference between magnetic dipole and electric quadrupole terms. We find that the extinction coefficient k defines the value of this imaginary part of the fraction in equation 9 and this is shown in Figure S11, section XI. \par

The origin of the magnetic response at these optical frequencies where the permeability, $\mu$, is 1 has been a subject of very active research \cite{Monticone2014}. Clearly, there is a strong high frequency magnetic response in our materials as is made clear through the presence of magnetic multipoles in our work. Landau and Lifshitz \cite{Landau1961} developed the argument that spatial dispersion of the refractive index results in a magnetic response and this argument has been expanded on by the work of Agranovich \cite{Agranovich2006}. Our thin film SiN sample featuring zero reflection with Fabry Perot and Brewster angle effects is exhibiting such spatial dispersion. Like the work of \cite{Li2021}, a magnetic response is demonstrated in thin films.\par

\bibliography{Bib}% Produces the bibliography via BibTeX.

\end{document}